\documentclass[aps,pre,showkeys,showpacs,preprint,longbibliography]{revtex4-1}

\usepackage{hyperref}
\usepackage{pgf,tikz}
\usetikzlibrary{arrows, positioning, calc, patterns}
\usepackage{amsmath,amssymb,amsthm}
\usepackage{graphicx}
\usepackage{xcolor}
\usepackage{xparse}
\usepackage{mathrsfs}

\theoremstyle{definition}

\newcommand{\mc}[1]{\ensuremath{\mathcal{#1}}}

\newcommand{\mbf}[1]{\ensuremath{\mathbf{#1}}}

\renewcommand{\d}{\ensuremath{\mathrm{d}}}
\DeclareDocumentCommand{\D}{omm}{\ensuremath{%
\IfNoValueTF{#1}{\frac{\d #2}{\mathrm{d} #3}}{\frac{\d^{#1} #2}{\d {#3}^{#1}}}}}
\DeclareDocumentCommand{\P}{omm}{\ensuremath{%
\IfNoValueTF{#1}{\frac{\partial #2}{\partial #3}}{\frac{\partial^{#1} #2}{\partial {#3}^{#1}}}}}
\DeclareDocumentCommand{\Pin}{omm}{\ensuremath{%
\IfNoValueTF{#1}{{\partial #2}/{\partial #3}}{{\partial^{#1} #2}/{\partial {#3}^{#1}}}}}

\DeclareDocumentCommand{\dx}{o}{\ensuremath{%
\IfNoValueTF{#1}{\Delta x}{\Delta x_{#1}}}}
\newcommand{\dt}{\ensuremath{\Delta t}}

\newcommand{\abs}[1]{\ensuremath{\vert #1 \vert}}

\newcommand{\prt}[1]{\ensuremath{\left( #1 \right) }}
\newcommand{\sbr}[1]{\ensuremath{\left[ #1 \right] }}
\newcommand{\bra}[1]{\ensuremath{\left\{ #1 \right\} }}

\newcommand{\set}[2]{\ensuremath{\left\{ #1 \mid #2 \right\}}}

\DeclareDocumentCommand{\seq}{oom}{\ensuremath{%
\IfNoValueTF{#1}{\bra{#3}_{i=1}^{\infty}}%
{\IfNoValueTF{#2}{\bra{#3}_{i=1}^{#1}}%
{\bra{#3}_{#1=1}^{#2}}}}}
\DeclareDocumentCommand{\isum}{oo}{\ensuremath{%
\IfNoValueTF{#1}{\IfNoValueTF{#2}{\sum_{i=1}^{\infty}}%
{\sum_{i=1}^{#2}}}%
{\sum_{#1=1}^{#2}}}}
\DeclareDocumentCommand{\icap}{oo}{\ensuremath{%
\IfNoValueTF{#1}{\IfNoValueTF{#2}{\bigcap_{i=1}^{\infty}}%
{\bigcap_{i=1}^{#2}}}%
{\bigcap_{#1=1}^{#2}}}}
\DeclareDocumentCommand{\icup}{oo}{\ensuremath{%
\IfNoValueTF{#1}{\IfNoValueTF{#2}{\bigcup_{i=1}^{\infty}}%
{\bigcup_{i=1}^{#2}}}%
{\bigcup_{#1=1}^{#2}}}}

\newcommand{\minus}{%
  \setbox0=\hbox{-}%
  \vcenter{%
    \hrule width\wd0 height \the\fontdimen8\textfont3%
  }%
}

\newcommand{\ei}{\mbf{e}_i}
\newcommand{\ec}[1]{\mbf{e}_{#1}}
\newcommand{\cl}{\mbf{\Pi}}
\newcommand{\cli}{\Pi_i}
\newcommand{\sumd}{\sum_{i=0}^N}
\newcommand{\ff}{\mbf{f}}

\newcommand{\uu}{\mbf{u}}
\newcommand{\feq}{\mbf{f}^{\text{eq}}}
\newcommand{\feqi}{f_i^{\text{eq}}}

\newcommand{\visc}[1]{\mu}
\newcommand{\cha}[1]{\mbf{c}_{#1}}
\newcommand{\strs}[1]{\gamma_{#1}}

\renewcommand{\dx}{\ensuremath{\Delta x}}
\newcommand{\xx}{\ensuremath{\mbf{x}}}
\newcommand{\Tau}{\mathcal{T}}
\newcommand{\diag}[1]{diag\prt{#1}}
\newcommand{\sbs}[1]{\mc{S}_{#1}}

\begin{document}

\title{Linear Analysis on Multiple-relaxation-time Lattice Boltzmann Method}

\author{Yu \surname{Wang}}
\affiliation{Department of Mechanical Engineering, University of Illinois at Urbana-Champaign, Urbana, IL 61801, USA}
\email{yuwang8@illinois.edu}
\thanks{corresponding author}

\author{Jingjing \surname{Shi}}
\affiliation{Department of Mechanical Engineering, Purdue University, West Lafayette, IN 47907, USA}
\email{shi153@purdue.edu}

\date{\today}

\begin{abstract}
The development of multiple-relaxation-time (MRT) Lattice Boltzmann method (LBM) is a significant contribution in improving the numerical behavior, revealing the math and physics mechanism and extending the application of LBM. However, some of the MRT schemes proposed previously are not physically-consistent. In this work, we take D2Q9 as a example to show how to derive physically-consistent MRT-LBM schemes by eigenvalue decomposition of the collision operator. In addition, the scheme is validated by the equivalence to Navier-Stokes equations and numerical simulations.
\end{abstract}



\maketitle

\section{Introduction}

The past two decades have seen the rapid growth of Lattice Boltzmann method (LBM)~\cite{chen_lattice_1998}~\cite{aidun_lattice-boltzmann_2010}. Among the many contributions, the proposal of multiple-relaxation-time (MRT) collision model takes its place. The significance of MRT-LBM is threefold: it improved the numerical behavior of LBM~\cite{dhumieres_multiplerelaxationtime_2002}, reveals the math and physics behind LBM~\cite{shan_general_2007}~\cite{lallemand_theory_2000}, and facilitate the extension of LBM~\cite{chai_multiple-relaxation-time_2011}~\cite{mccracken_multiple-relaxation-time_2005}~\cite{chen_multiple-relaxation-time_2010}.

Roughly speaking, the MRT collision model is an extension of the BGK collision model by decomposing the collision process into different modes and assigning different parameters for each mode. Obviously, the modes with different relaxation times should be independent, otherwise, inconsistency will happen. However, this rule is not always obeyed in practice. In this work, we will take D2Q9 lattice as an example to show how to get physically-consistent MRT schemes.

The rest of the paper will be organized as follows: in Section~\ref{LBM}, a brief introduction on LBM will be given with an emphasis on MRT collision model; in Section~\ref{theory}, we will propose a physically-consistent MRT scheme based on an eigenvalue decomposition on the linear approximation of the BGK collision operator; in Section~\ref{sec:main}, we will prove that the scheme reduces to Navier-Stokes equations at the macroscopic level; and the scheme will be further validated by simulation in Section~\ref{simu}; finally, we will conclude the work by Section~\ref{conc}.

\section{Lattice Boltzmann Method\label{LBM}} 

In this section, a brief introduction on Lattice Boltzmann method (LBM) will be given. 
Throughout the paper, scalars, vectors and tensors are denoted by lowercase letters, lowercase letters in boldface and uppercase letters, respectively. 

In LBM,  the flow of fluids is simulated by particles hopping on a lattice. Fixing the time step $\dt$ and the unit length of the lattice $\dx$, the velocity of the particles can only be chosen from a finite set of vectors 
$\set{\ei}{i=1,\ldots,N}$. 
Let $\ff(\xx, t)$ be the \emph{particle velocity distribution function} whose $i$-th component $f_i(\xx, t)$ gives the portion of particles with velocity $\ei$ at node $\xx$ at time $t$.
It satisfies the \emph{Lattice Boltzmann Equation} (LBE)
\begin{equation} \label{eq:lbe}
f_i(\xx + \ei \dt, t + \dt) = f_i(\xx, t) + \cli (\ff(\xx, t))
\end{equation}
where $i = 1,2,\ldots,N$ and $\cli (\ff( x, t))$ is the $i$-th component of the \emph{collision operator} that describe the effect of collision of particles.
In coding, the computation of \eqref{eq:lbe} is usually divided into two steps:
\begin{itemize}
	\item \textbf{Collision}: 
	\begin{equation} \label{eq:str}
		f_i'(\xx, t) = f_i(\xx, t) + \cli (\ff(\xx, t)),
	\end{equation}
	\item \textbf{Streaming}: 
	\begin{equation} \label{eq:col}
		f_i(\xx + \ei \dt, t + \dt) = f_i'(\xx, t).
	\end{equation}
\end{itemize}

\begin{figure}
\begin{tikzpicture}[scale = 1.5]
\draw[->] (0,0) -- (1.5,0);
\draw[->] (0,0) -- (0,1.5);
\node[right] at (1.5,0) {$x$};
\node[above] at (0,1.5) {$y$};

\draw (-1,1) -- (1,1);
\draw (-1,1) -- (-1,-1);
\draw (-1,-1) -- (1,-1);
\draw (1,-1) -- (1,1);

\draw[dashed] (-0.5,0.5) -- (0.5,0.5);
\draw[dashed] (-0.5,0.5) -- (-0.5,-0.5);
\draw[dashed] (-0.5,-0.5) -- (0.5,-0.5);
\draw[dashed] (0.5,-0.5) -- (0.5,0.5);

\draw[->,thick] (0,0) -- (1,0);
\draw[->,thick] (0,0) -- (0,1);
\draw[->,thick] (0,0) -- (-1,0);
\draw[->,thick] (0,0) -- (0,-1);
\draw[->,thick] (0,0) -- (1,1);
\draw[->,thick] (0,0) -- (-1,1);
\draw[->,thick] (0,0) -- (-1,-1);
\draw[->,thick] (0,0) -- (1,-1);

\node[above right] at (0,0) {$\ec{0}$};
\node[above right] at (1,0) {$\ec{1}$};
\node[above right] at (0,1) {$\ec{2}$};
\node[above left] at (-1,0) {$\ec{3}$};
\node[below right] at (0,-1) {$\ec{4}$};
\node[right] at (1,1) {$\ec{5}$};
\node[left] at (-1,1) {$\ec{6}$};
\node[left] at (-1,-1) {$\ec{7}$};
\node[right] at (1,-1) {$\ec{8}$};
\end{tikzpicture}
\caption{D2Q9 Scheme \label{fig1}}
\end{figure}
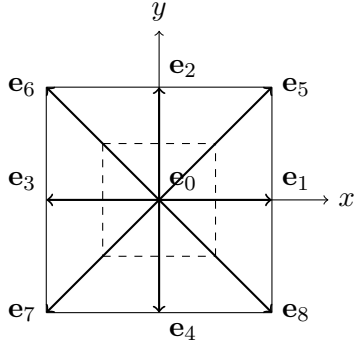

In this work, we will take the D2Q9 lattice (Figure~\ref{fig1}) as an example, where the set of velocity is taken to be
\[
\begin{array}{l l}
	\ec{0} = \sbr{0,0}, & \ec{1} = \sbr{ \kappa,0}, \\
	\ec{2} = \sbr{0, \kappa}, & \ec{3} = \sbr{\minus \kappa,0}, \\
	\ec{4} = \sbr{0,0}, & \ec{5} = \sbr{ \kappa, \kappa}, \\
	\ec{6} = \sbr{\minus \kappa, \kappa}, & \ec{7} = \sbr{\minus \kappa,\minus \kappa}, \\
	\ec{8} = \sbr{ \kappa,\minus \kappa}.
\end{array}
\]
where $\kappa = \dx/\dt$ is the characteristic velocity of the lattice.
The arguments in the rest of the paper extend easily to other kinds of lattices.

The particle velocity distribution function $\ff(\xx, t)$ is related to the macroscopic variables by 
\begin{align} \label{eq:rl}
\rho = \sumd f_i , \quad \rho \uu = \sumd f_i \ei,
\end{align}
where $\rho$ and $\uu$ are the macroscopic density and velocity respectively.
Due to the conservation of mass the momentum, for any  $\ff(\xx, t)$, the collision operator satisfies 
\begin{equation} \label{eq:cc}
\sumd \cli(\ff(\xx, t)) = 0,\quad \sumd \cli(\ff(\xx, t)) \ei = 0.
\end{equation}

In general, the particle velocity distribution function $\ff(\xx, t)$ cannot be deduced from the macroscopic quantities $\rho$ and $\uu$. But in equilibrium, the particle velocity distribution function $\feq(\xx,t)$ only depends on $\rho$ and $\uu$. In the D2Q9 lattice, by fitting $\ff(\xx, t)$ to the continuous equilibrium distribution function, we obtain that~\cite{qian_lattice_1992}
\begin{equation} \label{feq}
\feqi = \rho w_i \prt{1 + 3 \uu \cdot \ei + \frac{9}{2} \prt{\uu \cdot \ei}^2 - \frac{3}{2} \abs{\uu}^2}
\end{equation}
with $w_0 = 4/9$, $w_1 = w_3 = w_5 = w_7 = 1/9$ and $w_1 = w_3 = w_5 = w_7 = 1/36$.

Generally speaking, the collision operator is trying to restore the particle velocity distribution function $\ff(\xx, t)$ to its equilibrium distribution $\feq(\xx,t)$.
In the BGK collision model, the collision operator is taken to be a linear relaxation operator
\begin{equation}
	 \cl(\ff(\xx, t)) = \frac{\feq(\xx,t) - \ff(\xx, t)}{\tau_{BGK}}
\end{equation}
where the relaxation time $\tau_{BGK}$ is determined by kinematic viscosity $\nu$ by
\begin{equation}
	\tau = \frac{1}{2} + \frac{3 \nu \dt}{(\dx)^2}.
\end{equation}

In the multiple-relaxation-time collision model, the particle velocity distribution function $\ff(\xx, t)$ is decomposed into $N$ (the dimension of $\ff(\xx, t)$) modes and different relaxation times are assigned to each mode, i.e.
\begin{equation}
	\cl(\ff(\xx, t)) = P^{-1} \Tau^{-1} P(\feq(\xx,t) - \ff(\xx, t)),
\end{equation}
where $P$ is an invertible matrix and $\Tau = \diag{\tau_1, \tau_2, \ldots, \tau_N}$. When $\tau_1 = \ldots = \tau_N = \tau_{BGK}$, the MRT model degenerates to the BGK model.

The decomposition of $\ff(\xx, t)$ is not arbitrary; the modes corresponding to different relaxation times should be independent.
Previously, the most common decomposition scheme is given by~\cite{lallemand_theory_2000}
\begin{equation}
	P = 
	\begin{bmatrix}
		1 & 1 & 1 & 1 & 1 & 1 & 1 & 1 & 1 \\
		\minus4 & \minus1 & \minus1 & \minus1 & \minus1 & 2 & 2 & 2 & 2 \\
		4 & 2 & 2 & 2 & 2 & 1 & 1 & 1 & 1 \\
		0 & 1 & 0 & -1 & 0 & 1 & -1 & -1 & 1 \\
		0 & -2 & 0 & 2 & 0 & 1 & -1 & -1 & 1 \\
		0 & 0 & 1 & 0 & -1 & 1 & 1 & -1 & -1 \\
		0 & 0 & -2 & 0 & 2 & 1 & 1 & -1 & -1 \\
		0 & 1 & -1 & 1 & -1 & 0 & 0 & 0 & 0 \\
		0 & 0 & 0 & 0 & 0 & 1 & -1 & 1 & -1 \\
	\end{bmatrix},
\end{equation}
where the rows correspond respectively to density, energy, energy square, $x$-momentum, $x$-energy flux,  $y$-momentum, $y$-energy flux, diagonal component and off-diagonal component of stress tensor. However, the modes in this decomposition are not independent, e.g. ``energy'' and ``energy square''. Thus, inconsistency may happen when different relaxation times are assigned to them.

\section{Eigenvalue Decomposition \label{theory}} 
Before proposing a physically-consistent MRT scheme, we will first perform an eigenvalue analysis on the BGK collision operator.
The LBM holds when the macroscopic velocity $\abs{\uu(\xx,t)} \ll \dx/\dt$, i.e. the Mach number of the flow with respect to the characteristic velocity $\kappa$ is small. Therefore, \eqref{feq} is well approximated by its linearization
\begin{equation}
\feqi = \rho w_i \prt{1 + 3 \uu \cdot \ei}.
\end{equation}
In this case, $\feq$ depends linearly on $\ff$ by
\begin{equation} \label{eq:linlbm}
	\ff = T \feq
\end{equation}
where
\begin{equation}
T =
\begin{bmatrix}
	\frac{4}{9} & \frac{4}{9} & \frac{4}{9} & \frac{4}{9} & \frac{4}{9} & \frac{4}{9} & \frac{4}{9} & \frac{4}{9} & \frac{4}{9} \\
	\frac{1}{9} & \frac{4}{9} & \frac{1}{9} & \minus\frac{2}{9} & \frac{1}{9} & \frac{4}{9} & \minus\frac{2}{9} & \minus\frac{2}{9} & \frac{4}{9} \\
	\frac{1}{9} & \frac{1}{9} & \frac{4}{9} & \frac{1}{9} & \minus\frac{2}{9} & \frac{4}{9} & \frac{4}{9} & \minus\frac{2}{9} & \minus\frac{2}{9} \\
	\frac{1}{9} & \minus\frac{2}{9} & \frac{1}{9} & \frac{4}{9} & \frac{1}{9} & \minus\frac{2}{9} & \frac{4}{9} & \frac{4}{9} & \minus\frac{2}{9} \\
	\frac{1}{9} & \frac{1}{9} & \minus\frac{2}{9} & \frac{1}{9} & \frac{4}{9} & \minus\frac{2}{9} & \minus\frac{2}{9} & \frac{4}{9} & \frac{4}{9} \\
	\frac{1}{36} & \frac{1}{9} & \frac{1}{9} & \minus\frac{1}{18} & \minus\frac{1}{18} & \frac{7}{36} & \frac{1}{36} & \minus\frac{5}{36} & \frac{1}{36} \\
	\frac{1}{36} & \minus\frac{1}{18} & \frac{1}{9} & \frac{1}{9} & \minus\frac{1}{18} & \frac{1}{36} & \frac{7}{36} & \frac{1}{36} & \minus\frac{5}{36} \\
	\frac{1}{36} & \minus\frac{1}{18} & \minus\frac{1}{18} & \frac{1}{9} & \frac{1}{9} & \minus\frac{5}{36} & \frac{1}{36} & \frac{7}{36} & \frac{1}{36} \\
	\frac{1}{36} & \frac{1}{9} & \minus\frac{1}{18} & \minus\frac{1}{18} & \frac{1}{9} & \frac{1}{36} & \minus\frac{5}{36} & \frac{1}{36} & \frac{7}{36} 
\end{bmatrix}
\end{equation}
The set of eigenvalues of matrix $T$ are $\bra{1,1,1,0,0,0,0,0,0}$ and the corresponding eigenvectors are taken to be 
\begin{equation}
	\begin{split}
		\cha{1} & = \sbr{1, 1, 1, 1, 1, 1, 1, 1, 1}, \\
		\cha{2} & = \sbr{0, 1, 0, \minus1, 0, 1, \minus1, \minus1, 1}, \\
		\cha{3} & = \sbr{0, 0, 1, 0, \minus1, 1, 1, \minus1, \minus1}. \\
		\cha{4} & = \sbr{\minus\frac{1}{3}, \frac{2}{3}, \minus\frac{1}{3}, \frac{2}{3}, \minus\frac{1}{3}, \frac{2}{3}, \frac{2}{3}, \frac{2}{3}, \frac{2}{3}}, \\
		\cha{5} & = \sbr{0, 0, 0, 0, 0, 1, \minus1, 1, \minus1}. \\
		\cha{6} & = \sbr{\minus\frac{1}{3}, \minus\frac{1}{3}, \frac{2}{3}, \minus\frac{1}{3}, \frac{2}{3}, \frac{2}{3}, \frac{2}{3}, \frac{2}{3}, \frac{2}{3}}, \\
		\cha{7} & = \sbr{0, 1, 0, \minus1, 0, \minus2, 2, 2, \minus2}, \\
		\cha{8} & = \sbr{0, 0, 1, 0, \minus1, \minus2, \minus2, 2, 2}, \\
		\cha{9} & = \sbr{\minus1, 0, 0, 0, 0, 4, 4, 4, 4}.
	\end{split}
\end{equation}
The nine eigenvectors, each representing a physical mode, are divided into three groups: $\bra{\cha{1}, \cha{2}, \cha{3}}$, $\bra{\cha{4}, \cha{5}, \cha{6}}$ and $\bra{\cha{7}, \cha{8}, \cha{9}}$.

In the first group, $\cha{1}, \cha{2}, \cha{3}$ correspond to density, $x$-momentum and $y$-momentum respectively,
\begin{equation}
	\begin{split}
		& \rho(\xx,t) = \cha{1} \cdot \ff(\xx, t) \\
		& \rho(\xx,t) u_x(\xx,t) = \cha{2} \cdot \ff(\xx, t) \\
		& \rho(\xx,t) u_y(\xx,t) = \cha{3} \cdot \ff(\xx, t) \\
	\end{split}
\end{equation}
By~\eqref{eq:str} and~\eqref{eq:linlbm}, for each $k = 1,2,3$ and any $\tau_k$, we have
\begin{equation}
	\cha{k} \cdot \ff'(\xx, t) = \cha{k} \cdot \ff(\xx, t).
\end{equation}
Therefore, mass and momentum are conservative in the \textbf{collision} step.

In the second group, $\cha{4},\cha{5},\cha{6}$ correspond to the components of the stress tensor $\Gamma$~\cite{chen_lattice_1998} by
\begin{equation} \label{eq:gam}
	\begin{split}
		& \strs{xx}(\xx,t)/\nu = \cha{4} \cdot \ff(\xx, t) \\
		& \strs{xy}(\xx,t)/\nu = \cha{5} \cdot \ff(\xx, t) \\
		& \strs{yy}(\xx,t)/\nu = \cha{6} \cdot \ff(\xx, t) \\
	\end{split}
\end{equation}
where
\begin{equation}
	\Gamma = \begin{bmatrix}
		\strs{xx} & \strs{xy} \\
		\strs{xy} & \strs{yy}
	\end{bmatrix}
\end{equation}
The physical picture is explained in the following way.
Take the control volume as shown by the dashed line in Figure~\ref{fig1}, the flux of $x$-momentum in the $x$-direction crossing the boundary of $\Gamma$ is given by
\begin{equation}
	\Phi_{xx} = \sbr{0,1,0,1,0,1,1,1,1} \cdot \ff(\xx, t)
\end{equation}
Similarly, the flux of $y$-momentum in the $y$-direction crossing the boundary of $\Gamma$ is given by
\begin{equation}
	\Phi_{yy} = \sbr{0,0,1,0,1,1,1,1,1} \cdot \ff(\xx, t).
\end{equation}
The flux of $x$-momentum in the $y$-direction crossing the boundary of $\Gamma$ is equal to flux of $y$-momentum in the $x$-direction, that is,
\begin{equation}
	\Phi_{xy} = \Phi_{yx} = \sbr{0,0,0,0,1,\minus1,1,\minus1} \cdot \ff(\xx, t).
\end{equation}
Noting that
\begin{equation}
	\begin{bmatrix}
		\Phi_{xx} & \Phi_{xy} \\
		\Phi_{yx} & \Phi_{yy}
	\end{bmatrix}
	= \Gamma/\nu
	+ p \mbf{I}
\end{equation}
and $p = \frac{\rho}{3}$, we obtain~\eqref{eq:gam}. 

As shown above, the three eigenvalues in this group correspond to the transportation of momentum due to viscosity. Therefore, the relaxation times should be taken as $\tau_4 = \tau_5 = \tau_6 = \tau_{BGK}$.
By~\eqref{eq:str} and~\eqref{eq:linlbm}, 
for each $k = 4,5,6$, we have
\begin{equation}
	\cha{k} \cdot \ff'(\xx, t) = (1 - \frac{1}{\tau_{BGK}}) \ \cha{k} \cdot \ff(\xx, t),
\end{equation}
in the \textbf{collision} step.

Finally, in the third group, $\cha{7},\cha{8},\cha{9}$ represent no macroscopic physical quantity, namely, they are redundant degrees of freedom in computation. Therefore, the relaxation times $\tau_7, \tau_8, \tau_9$ can be chosen arbitrarily.  By~\eqref{eq:str} and~\eqref{eq:linlbm}, 
for each $k = 7,8,9$, we have
\begin{equation}
	\cha{k} \cdot \ff'(\xx, t) = (1 - \frac{1}{\tau_{k}}) \ \cha{k} \cdot \ff(\xx, t),
\end{equation}
in the \textbf{collision} step.

Based on the eigenvalue decomposition, we propose the following MRT collision operator
\begin{equation} \label{eq:mrt}
	\cl(\ff(\xx, t)) = - Q^{-1} S Q \ \ff(\xx, t),
\end{equation}
where
\begin{equation}
	Q = \begin{bmatrix}
		\cha{1} \\ \cha{2} \\ \cha{3} \\ \cha{4} \\ \cha{5} \\ \cha{6} \\ \cha{7} \\ \cha{8} \\ \cha{9}
	\end{bmatrix} = 
	\begin{bmatrix}
		1 & 1 & 1 & 1 & 1 & 1 & 1 & 1 & 1 \\
		0 & 1 & 0 & \minus1 & 0 & 1 & \minus1 & \minus1 & 1 \\
		0 & 0 & 1 & 0 & \minus1 & 1 & 1 & \minus1 & \minus1 \\
		\minus\frac{1}{3} & \frac{2}{3} & \minus\frac{1}{3} & \frac{2}{3} & \minus\frac{1}{3} & \frac{2}{3} & \frac{2}{3} & \frac{2}{3} & \frac{2}{3} \\
		0 & 0 & 0 & 0 & 0 & 1 & \minus1 & 1 & \minus1 \\
		\minus\frac{1}{3} & \minus\frac{1}{3} & \frac{2}{3} & \minus\frac{1}{3} & \frac{2}{3} & \frac{2}{3} & \frac{2}{3} & \frac{2}{3} & \frac{2}{3} \\
		0 & 1 & 0 & \minus1 & 0 & \minus2 & 2 & 2 & \minus2 \\
		0 & 0 & 1 & 0 & \minus1 & \minus2 & \minus2 & 2 & 2 \\
		\minus1 & 0 & 0 & 0 & 0 & 4 & 4 & 4 & 4
	\end{bmatrix}
\end{equation}
\begin{equation}
	S = \diag{0, 0, 0, \frac{1}{\tau_{BGK}}, \frac{1}{\tau_{BGK}}, \frac{1}{\tau_{BGK}}, \frac{1}{\tau_7},\frac{1}{\tau_8}, \frac{1}{\tau_9}}
\end{equation}
The relaxation time $\tau_{BGK}$ is determined by kinetic viscosity $\nu$ by
\begin{equation}
	\tau = \frac{1}{2} + \frac{3 \nu \dt}{(\dx)^2}
\end{equation}
Obviously,
\begin{equation} \label{eq:rrl}
		\cha{k} \cdot Q^{-1} S Q = \begin{cases}
			0, &\text{ if } k = 1,2,3 \\
			\cha{k}/\tau_{BGK}, &\text{ if } k = 4,5,6 \\
			\cha{k}/\tau_{k}, &\text{ if } k = 7,8,9
		\end{cases}
\end{equation}

\section{Multiple-relaxation-time collision operator} \label{sec:main}
The MRT scheme proposed in Section~\ref{theory} is validated by the equivalence to Navier-Stokes equations through Chapman-Enskog expansion~\cite{frisch1987lattice}. Assuming that $\dx = \dt = \varepsilon$, then the Taylor expansion of LBE~\eqref{eq:lbe} gives
\begin{equation} \label{taylor}
	\varepsilon \prt{\P{f_i}{t} + \ei \cdot \nabla f_i} + \varepsilon^2 \prt{\frac{1}{2} \P[2]{f_i}{t} + \ei \cdot \nabla \P{f_i}{t} + \frac{1}{2} \ei \ei : \nabla \nabla f_i}
	= \cli
\end{equation}
Let $t_1$ be the convection time scale and $t_2$ be the diffusion time scale, then the time derivative decomposes to
\begin{equation} \label{eq:time}
	\P{}{t} = \P{}{t_1} + \varepsilon \P{}{t_2}.
\end{equation}
Accordingly, the distribution function decomposes near the equilibrium to
\begin{equation} \label{eq:distribution}
	f_i (\xx, t) = \feqi (\xx, t) + \varepsilon f^{(1)}_i (\xx, t) + \varepsilon^2 f^{(2)}_i (\xx, t) + \ldots.
\end{equation}
By conservation of mass and momentum~\eqref{eq:rl}, we have
\begin{equation} \label{eq:rl2}
	\sumd \feqi = \rho, \quad \sumd \feqi \ei = \rho \uu.
\end{equation}
and for $s = 1, 2, \ldots$
\begin{equation}  \label{eq:rl3}
	\sumd f_i^{(s)} = \sumd f_i^{(s)} \ei = 0.
\end{equation}

Plugging~\eqref{eq:time}\eqref{eq:distribution} into~\eqref{taylor} and noting that
\begin{equation}
	Q^{-1} S Q \ \feq = 0,
\end{equation}
we obtain that
\begin{equation} \label{eq:od1}
	\P{\feqi}{t_1} + \ei \cdot \nabla \feqi = - \sbr{Q^{-1} S Q \ \ff^{(1)}}_i
\end{equation}
to the order $\varepsilon$, and
\begin{equation} \label{eq:od2}
	\P{\feqi}{t_2} + \P{f_i^{(1)}}{t_1} + \ei \cdot \nabla f_i^{(1)} + \frac{1}{2} \P[2]{f_i^{(1)}}{t} + \ei \cdot \nabla \P{f_i^{(1)}}{t} + \frac{1}{2} \ei \ei : \nabla \nabla f_i^{(1)}
	 = - \sbr{Q^{-1} S Q \ \ff^{(2)}}_i
\end{equation}
to the order $\varepsilon^2$.
Using~\eqref{eq:od1}, \eqref{eq:od2} simplifies to
\begin{equation} \label{eq:od3}
	\P{\feqi}{t_2} + \prt{\P{\sbr{M \ \ff^{(1)}}_i}{t_1} + \ei \cdot \nabla \sbr{M \ \ff^{(1)}}_i} 
	 = - \sbr{Q^{-1} S Q \ \ff^{(2)}}_i	
\end{equation}
where $M = \mbf{I} - Q^{-1} S Q/2$.

Using~\eqref{eq:rl2}\eqref{eq:rl3}, the mass equation is given by $\sumd$ (\eqref{eq:od1} + \eqref{eq:od3}) as
\begin{equation} 
	\P{\rho}{t} + \nabla \cdot (\rho \uu) = 0
\end{equation}
and the momentum equation is given by $\sumd \ei \cdot $  (\eqref{eq:od1} + \eqref{eq:od3}) as
\begin{equation} \label{eq:mom}
	\P{(\rho \uu)}{t} + \nabla \cdot \sumd \prt{\ei \ei \feqi + \ei \ei \sbr{M \ \ff^{(1)}}_i } = 0.
\end{equation}
where 
\begin{equation}
	\sumd \ei \ei \feqi = p \mbf{I} + \rho \uu \uu.
\end{equation}
and by~\eqref{eq:rrl}
\begin{equation}
	\sumd \ei \ei \sbr{M \ \ff^{(1)}}_i = \prt{1 - \frac{1}{2 \tau_{BGK}}} \begin{bmatrix}
		\cha{4} \cdot \ff^{(1)} & \cha{5} \cdot \ff^{(1)} \\
		\cha{5} \cdot \ff^{(1)} & \cha{6} \cdot \ff^{(1)}
	\end{bmatrix}
	= \Gamma
\end{equation}

From the discussion above, we make the following remarks. Let $\sbs{1}, \sbs{2}, \sbs{3}$ be the linear subspaces spanned by  $\bra{\cha{1}, \cha{2}, \cha{3}}$, $\bra{\cha{4}, \cha{5}, \cha{6}}$ and $\bra{\cha{4}, \cha{5}, \cha{6}, \cha{7}, \cha{8}, \cha{9}}$ respectively, then
\begin{itemize}
	\item $\cha{1}, \cha{2}, \cha{3}$ is arbitrary as long as $\sbs{1}$ is preserved;
	\item $\cha{4}, \cha{5}, \cha{6}$ is arbitrary as long as $\sbs{2}$ is preserved;
	\item $\cha{7}, \cha{8}, \cha{9}$ is arbitrary as long as $\sbs{3}$ is preserved;
\end{itemize}
Though the choice of $\tau_7, \tau_8, \tau_9$ has no influence on the result, for stability, we require that
\begin{equation}
	\tau_7, \tau_8, \tau_9 \geq \frac{1}{2}.
\end{equation}

\section{Simulations \label{simu}}
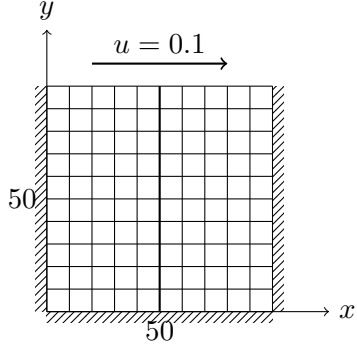
\begin{figure}
\begin{tikzpicture}[scale = 3]
\draw[->] (0,0) -- (1.25,0);
\draw[->] (0,0) -- (0,1.25);
\node[right] at (1.25,0) {$x$};
\node[above] at (0,1.25) {$y$};

\draw (0,0) -- node[below]{$50$} (1,0);
\draw (0,0) -- node[left]{$50$} (0,1);
\draw (0,1) -- (1,1);
\draw (1,0) -- (1,1);
\draw[->,thick] (0.2,1.1) -- node[above,midway]{$u = 0.1$} (0.8,1.1);

\fill[pattern=north east lines] (0,0) -- (0,1) -- (-0.05,1) -- (-0.05,0);
\fill[pattern=north east lines] (1,0) -- (1,1) -- (1.05,1) -- (1.05,0);
\fill[pattern=north east lines] (0,0) -- (1,0) -- (1,-0.05) -- (0,-0.05);

\draw (0.1,0) -- (0.1,1);
\draw (0.2,0) -- (0.2,1);
\draw (0.3,0) -- (0.3,1);
\draw (0.4,0) -- (0.4,1);
\draw[thick] (0.5,0) -- (0.5,1);
\draw (0.6,0) -- (0.6,1);
\draw (0.7,0) -- (0.7,1);
\draw (0.8,0) -- (0.8,1);
\draw (0.9,0) -- (0.9,1);

\draw (0,0.1) -- (1,0.1);
\draw (0,0.2) -- (1,0.2);
\draw (0,0.3) -- (1,0.3);
\draw (0,0.4) -- (1,0.4);
\draw (0,0.5) -- (1,0.5);
\draw (0,0.6) -- (1,0.6);
\draw (0,0.7) -- (1,0.7);
\draw (0,0.8) -- (1,0.8);
\draw (0,0.9) -- (1,0.9);

\end{tikzpicture}
\caption{Cavity flow\label{fig2}}
\end{figure}

In this section, the simulation results of 2-D cavity flow will be given to verify the MRT scheme proposed in Section~\ref{sec:main}. As shown in Figure~\ref{fig2}, the size of the domain is $50 \times 50$. The west boundary $x=0$, south boundary $y=0$ and east boundary $x=50$ are solid walls. The north boundary $y=50$ moves at a constant speed of $(0.1,0)$. 

In the simulations, the lattice has $51 \times 51$ nodes where the unit length $\dx = 1$ and the time step $\dt = 1$. Initially, the density and velocity on each node are set to be $\rho = 1$ and $(u,v) = (0,0)$ respectively and the velocity boundary conditions are applied at the boundary~\cite{zou_pressure_1995}.  The relaxation times are assigned differently in the following seven cases
\begin{itemize}
	\item Case 1: $S = \diag{0, 0, 0, \frac{1}{\tau_a}, \frac{1}{\tau_a}, \frac{1}{\tau_a}, \frac{1}{\tau_a},\frac{1}{\tau_a}, \frac{1}{\tau_a}}$ \\
	\item Case 2: $S = \diag{0, 0, 0, \frac{1}{\tau_b}, \frac{1}{\tau_a}, \frac{1}{\tau_a}, \frac{1}{\tau_a},\frac{1}{\tau_a}, \frac{1}{\tau_a}}$ \\
	\item Case 3: $S = \diag{0, 0, 0, \frac{1}{\tau_a}, \frac{1}{\tau_b}, \frac{1}{\tau_a}, \frac{1}{\tau_a},\frac{1}{\tau_a}, \frac{1}{\tau_a}}$ \\
	\item Case 4: $S = \diag{0, 0, 0, \frac{1}{\tau_a}, \frac{1}{\tau_a}, \frac{1}{\tau_b}, \frac{1}{\tau_a},\frac{1}{\tau_a}, \frac{1}{\tau_a}}$ \\
	\item Case 5: $S = \diag{0, 0, 0, \frac{1}{\tau_a}, \frac{1}{\tau_a}, \frac{1}{\tau_a}, \frac{1}{\tau_b},\frac{1}{\tau_a}, \frac{1}{\tau_a}}$ \\
	\item Case 6: $S = \diag{0, 0, 0, \frac{1}{\tau_a}, \frac{1}{\tau_a}, \frac{1}{\tau_a}, \frac{1}{\tau_a},\frac{1}{\tau_b}, \frac{1}{\tau_a}}$ \\
	\item Case 7: $S = \diag{0, 0, 0, \frac{1}{\tau_a}, \frac{1}{\tau_a}, \frac{1}{\tau_a}, \frac{1}{\tau_a},\frac{1}{\tau_a}, \frac{1}{\tau_b}}$
\end{itemize}
where $\nu_a = 0.2$, $\nu_b = 0.6$, $\tau_a = \frac{1}{2} + 3 \nu_1$ and $\tau_b = \frac{1}{2} + 3 \nu_2$.

\begin{figure}
\centering
\includegraphics[width = 0.45\textwidth]{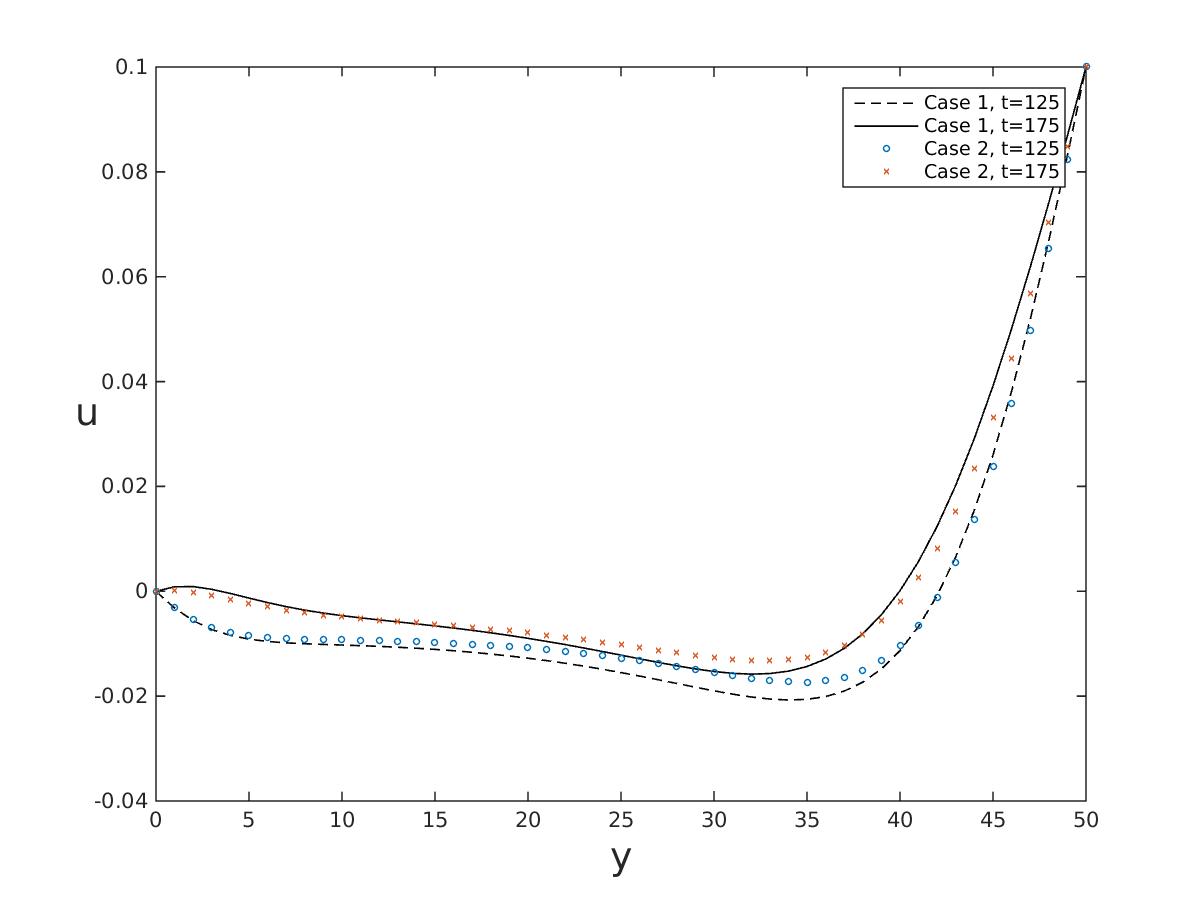}
\caption{Change the relaxation time of mode $\cha{4}$ \label{res1}}
\end{figure}

\begin{figure}
\centering
\includegraphics[width = 0.45\textwidth]{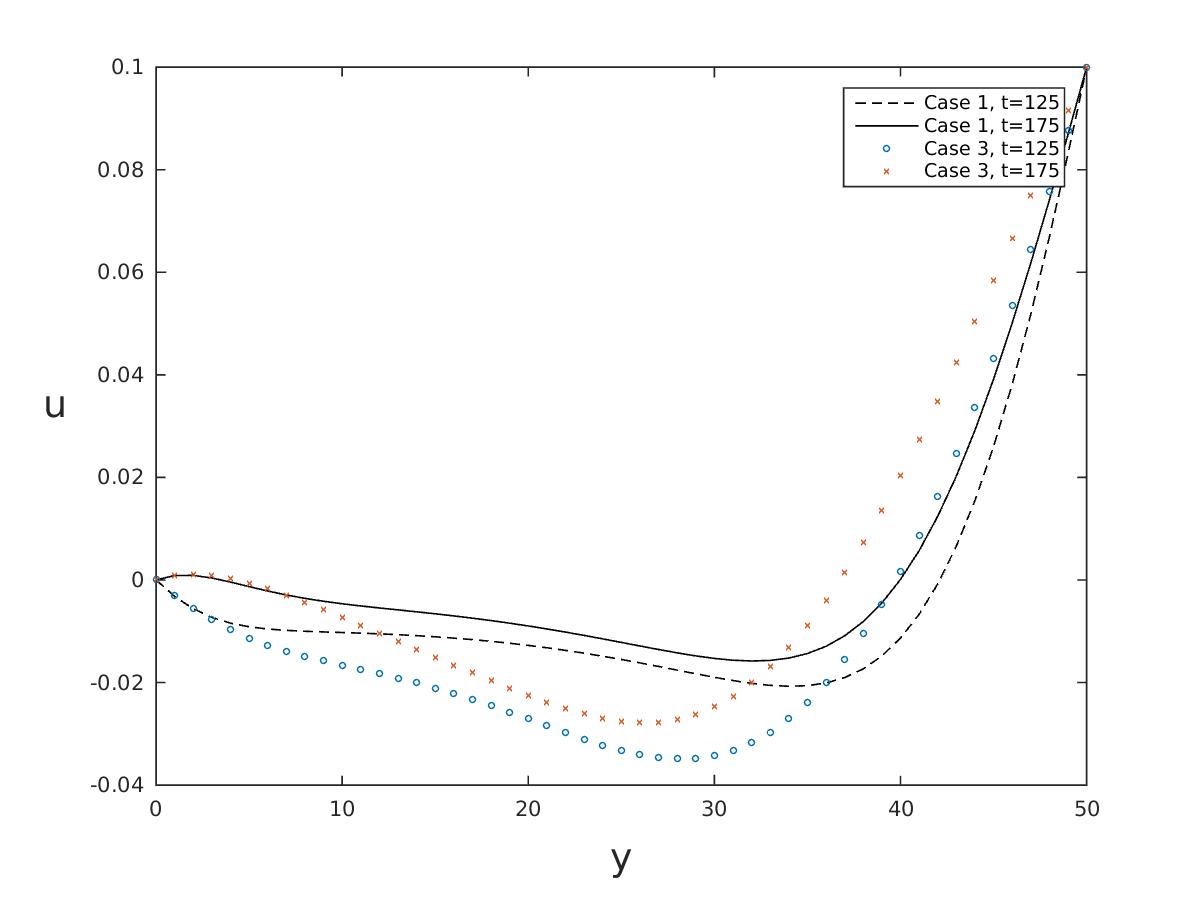}
\caption{Change the relaxation time of mode $\cha{5}$ \label{res2}}
\end{figure}

\begin{figure}
\centering
\includegraphics[width = 0.45\textwidth]{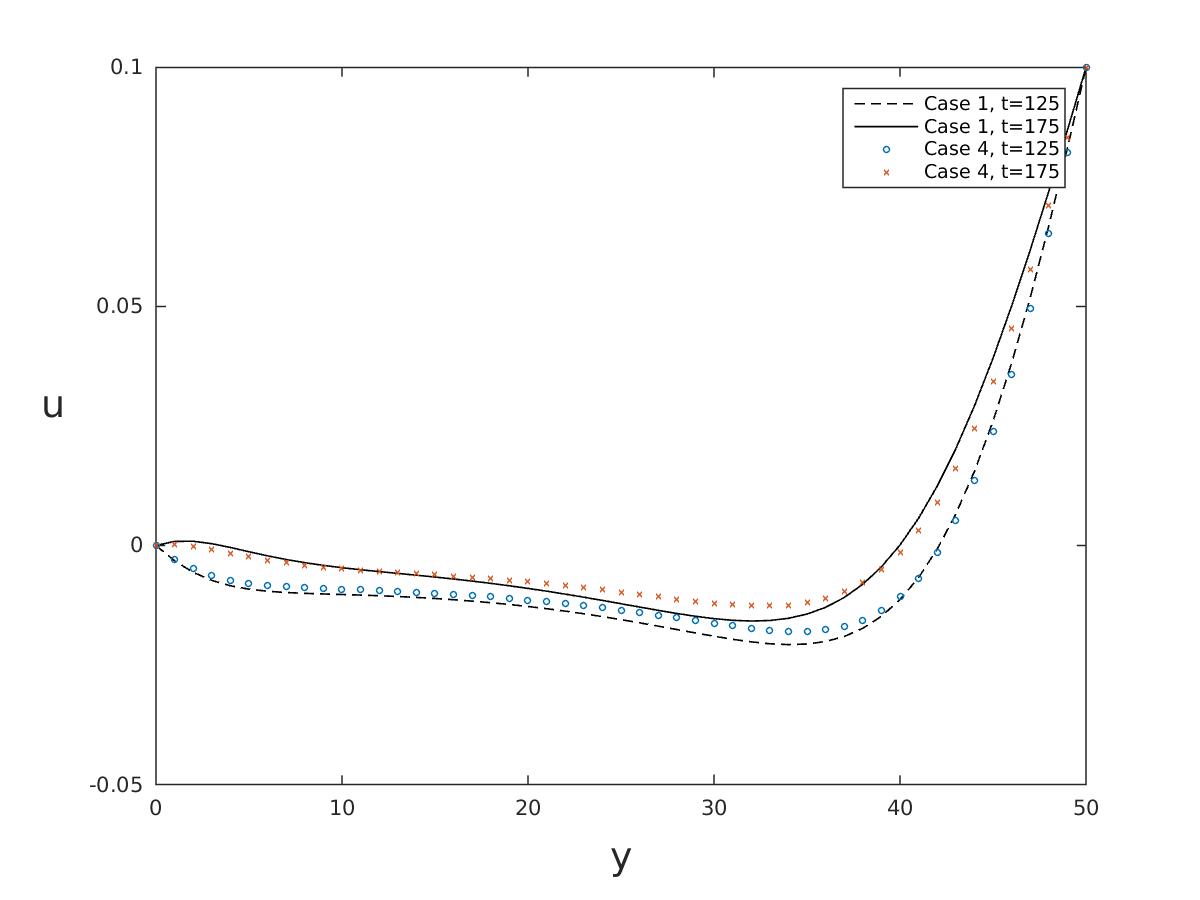}
\caption{Change the relaxation time of mode $\cha{6}$ \label{res3}}
\end{figure}

\begin{figure}
\centering
\includegraphics[width = 0.45\textwidth]{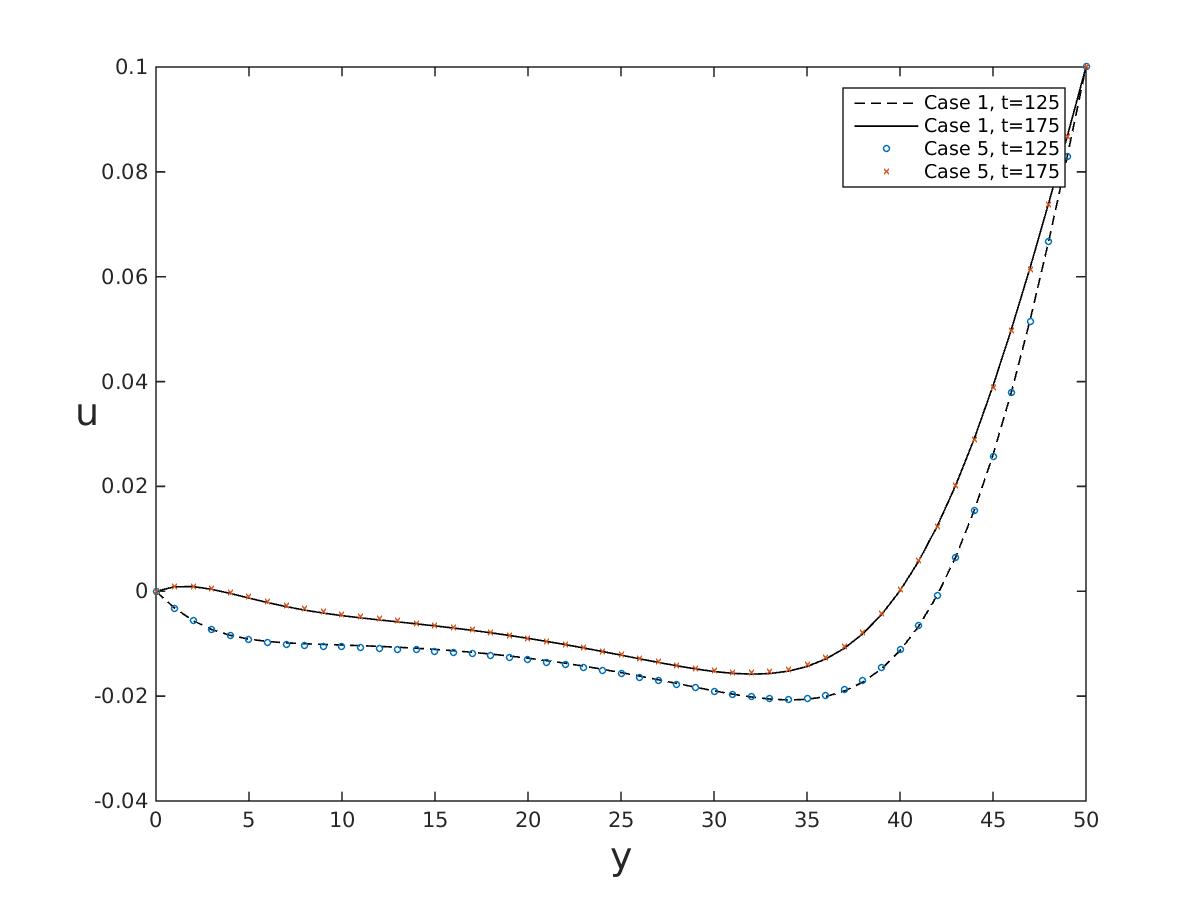}
\caption{Change the relaxation time of mode $\cha{7}$ \label{res4}}
\end{figure}

\begin{figure}
\centering
\includegraphics[width = 0.45\textwidth]{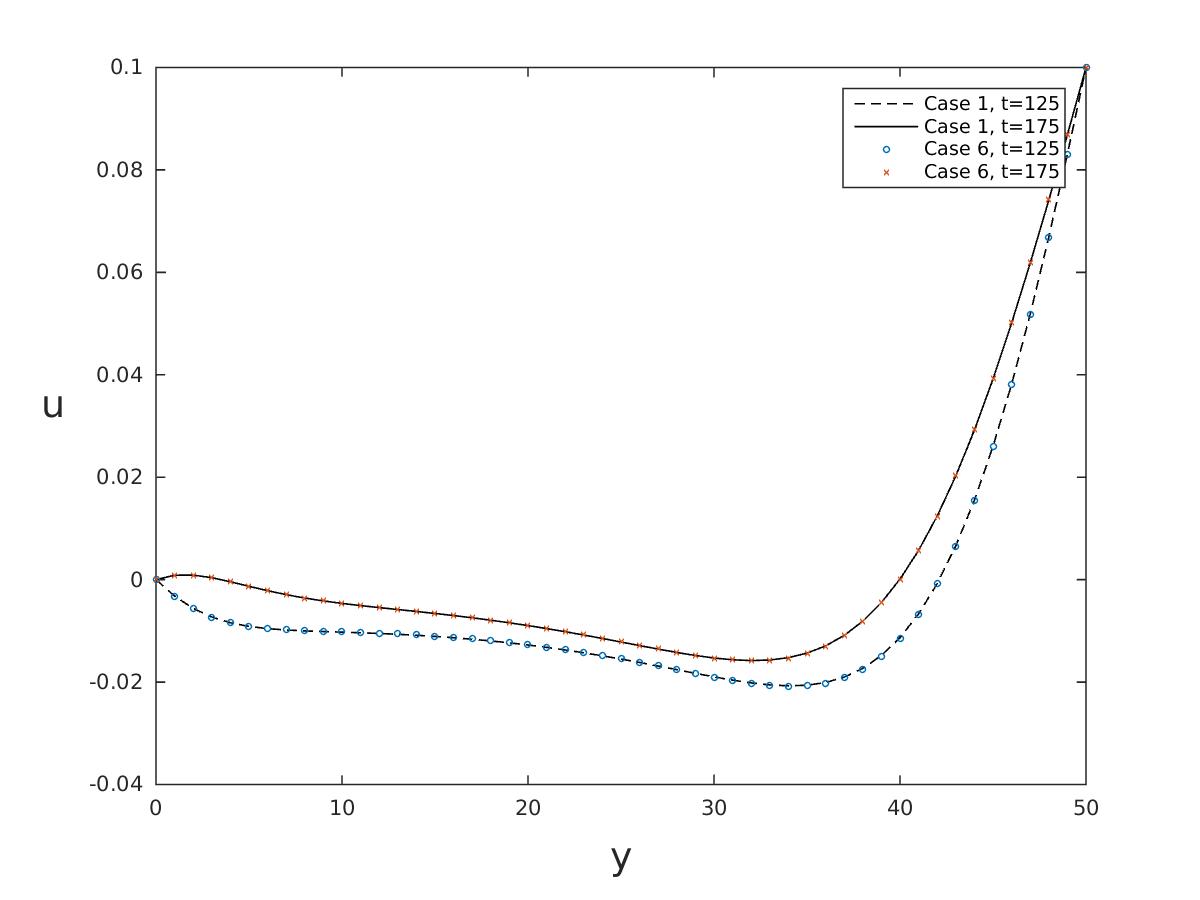}
\caption{Change the relaxation time of mode $\cha{8}$ \label{res5}}
\end{figure}

\begin{figure}
\centering
\includegraphics[width = 0.45\textwidth]{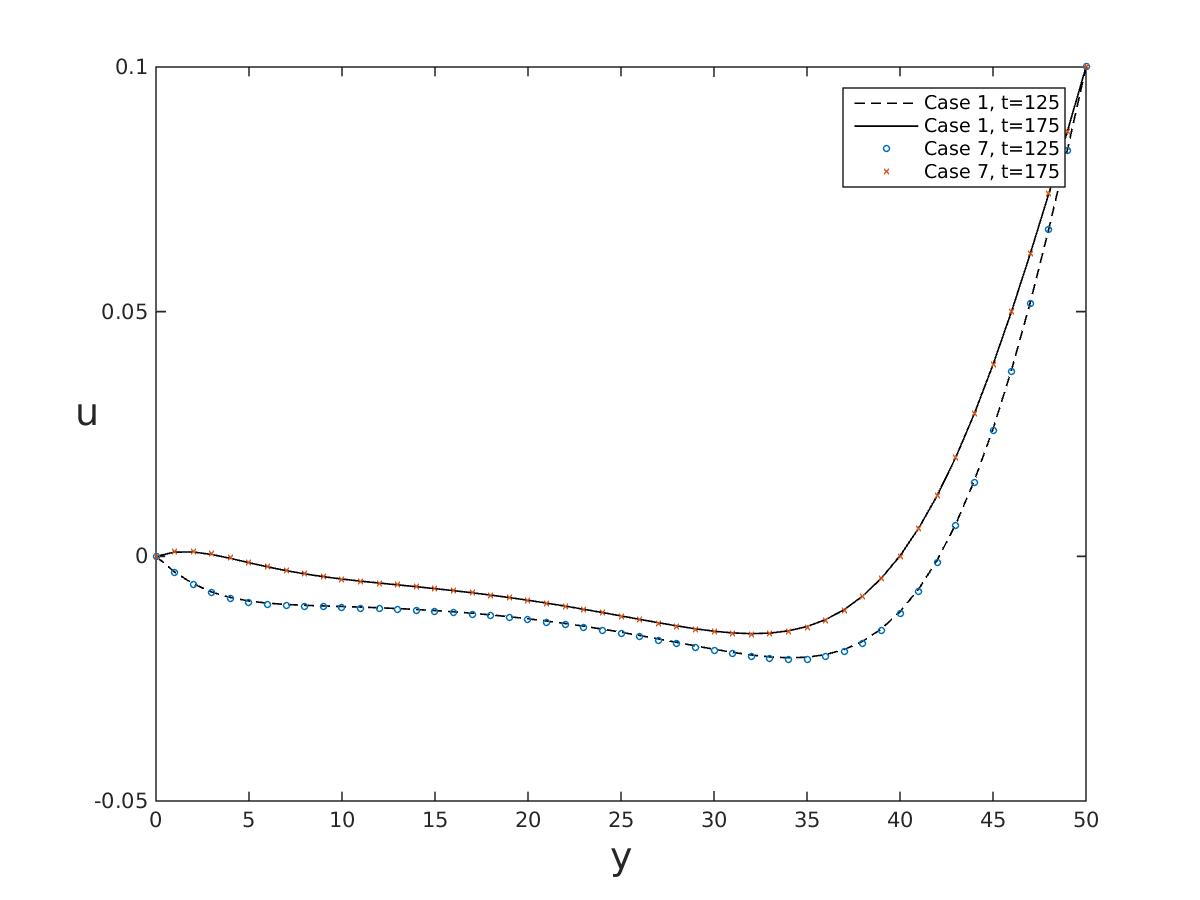}
\caption{Change the relaxation time of mode $\cha{9}$ \label{res6}}
\end{figure}

In Figure~\ref{res1} -- \ref{res6}, the plots of $x$-velocity against $y$ along the line $x = 25$ (shown by the thick solid line in Figure~\ref{fig2}) at time $t=125$ and $t=175$ are compared between different assignments of relaxation time.
The results show that changing the relaxation time of mode $\cha{4}$, $\cha{5}$ and $\cha{6}$ affects the computation results, while changing the relaxation time of mode $\cha{7}$, $\cha{8}$ and $\cha{9}$ does not. In addition, the relaxation time of mode $\cha{5}$ has a greater influence than the relaxation time of mode $\cha{4}$ and $\cha{6}$. This is because the shear stress plays a more important role than the normal stresses in this physical process.

\section{Conclusion\label{conc}}
In this work, we proposed a way of deriving physically-consistent MRT-LBM schemes based on eigenvalue decomposition of the collision operator. We showed that the scheme is equivalent to the Navier-Stokes equations at the macroscopic level and is in agreement with the simulation results.

\end{document}